\documentclass[aps,prl,twocolumn,showpacs,superscriptaddress]{revtex4}  % for review and submission
\usepackage{morefloats}
\usepackage{amsfonts}
\usepackage{amsmath}
\usepackage{amssymb}
\usepackage{graphicx}
\usepackage{caption}
\usepackage[singlelinecheck=off,skip=-8pt]{subcaption}
\begin{document}
\title{Monitoring superparamagnetic Langevin behavior of individual ${\rm SrRuO_3}$ nanostructures}

\author{Omer Sinwani}
\email{omersinwani@gmail.com}
\affiliation{Department of Physics, Nano-magnetism Research
Center, Institute of Nanotechnology and Advanced Materials,
Bar-Ilan University, Ramat-Gan 52900, Israel}
\author{James W. Reiner}
\affiliation{HGST, San Jose, California 95135, USA}

\author{Lior Klein}
\affiliation{Department of Physics, Nano-magnetism Research
Center, Institute of Nanotechnology and Advanced Materials,
Bar-Ilan University, Ramat-Gan 52900, Israel}


\begin{abstract}

Patterned nanostructures on the order of 200 nm $\times$ 200 nm of the itinerant ferromagnet ${\rm SrRuO_3}$ give rise to superparamagnetic behavior below the Curie temperature (${\rm \sim 150 \ K}$) down to a sample-dependent blocking temperature. We monitor the superparamagnetic fluctuations of an individual volume and demonstrate that the field dependence of the time-averaged magnetization is well described by the Langevin equation. On the other hand, the  rate of the fluctuations suggests that the volume in which the magnetization fluctuates is smaller by more than an order of magnitude. We suggest that switching via nucleation followed by propagation gives rise to Langevin behavior of the total  volume, whereas the switching rate is determined by a much smaller nucleation volume.

\end{abstract}

\maketitle

\begin{figure}[t]
\includegraphics[trim=70 30 0 0,scale=0.27,angle=-90]{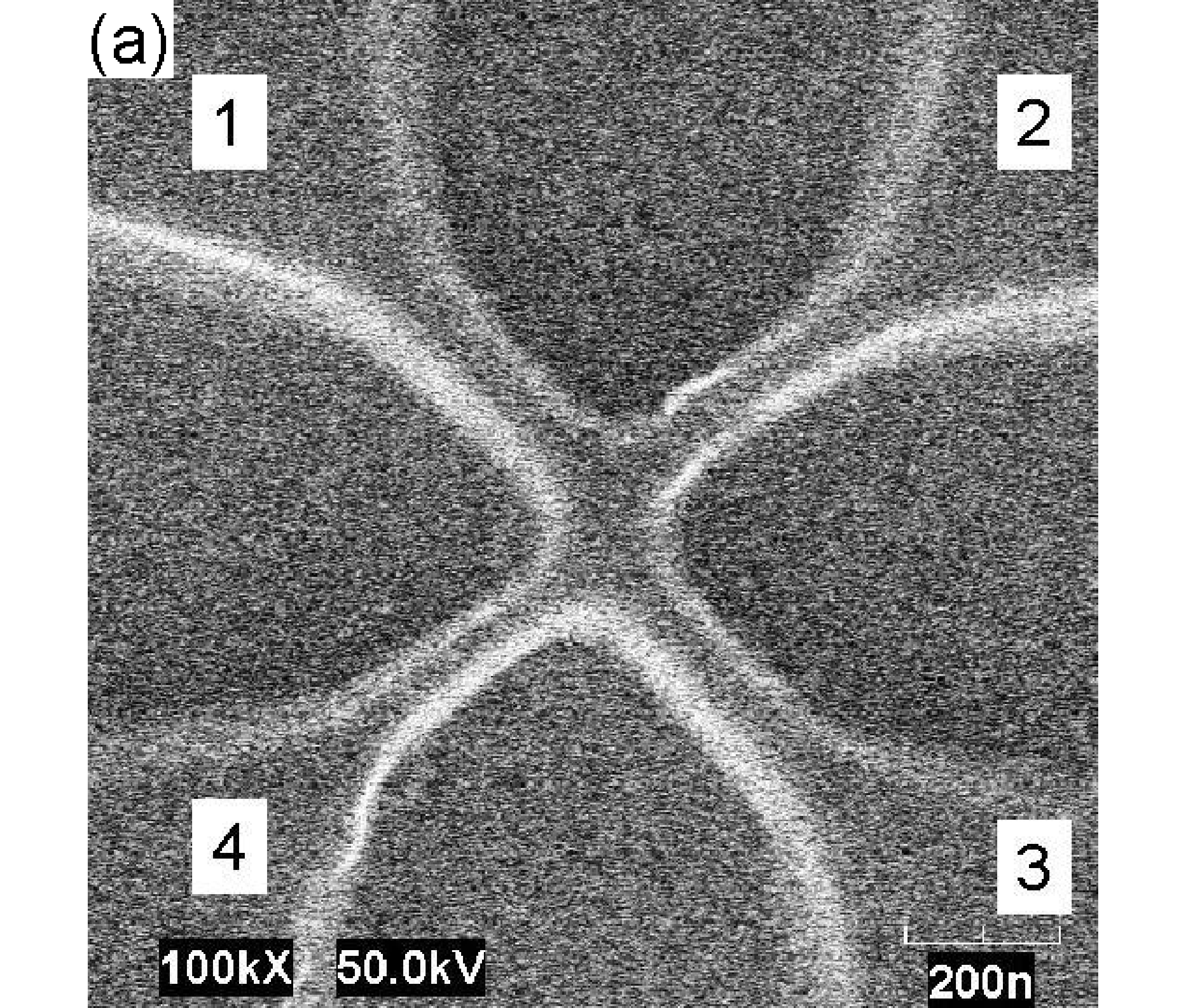}
\includegraphics[trim=50 0 40 0,scale=0.35,angle=-90]{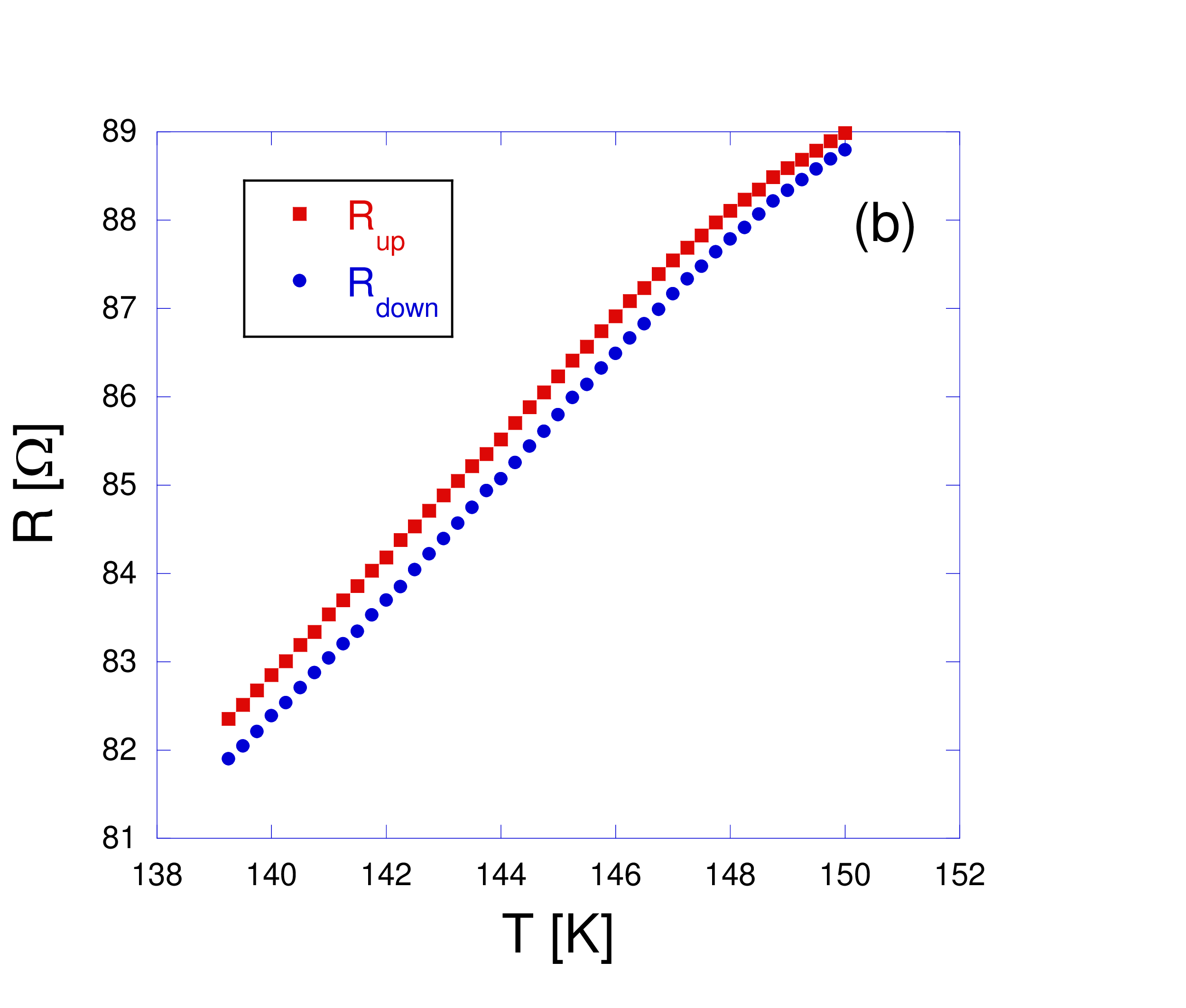}
\includegraphics[trim=0 0 100 0,keepaspectratio=true,scale=0.35,angle=-90]{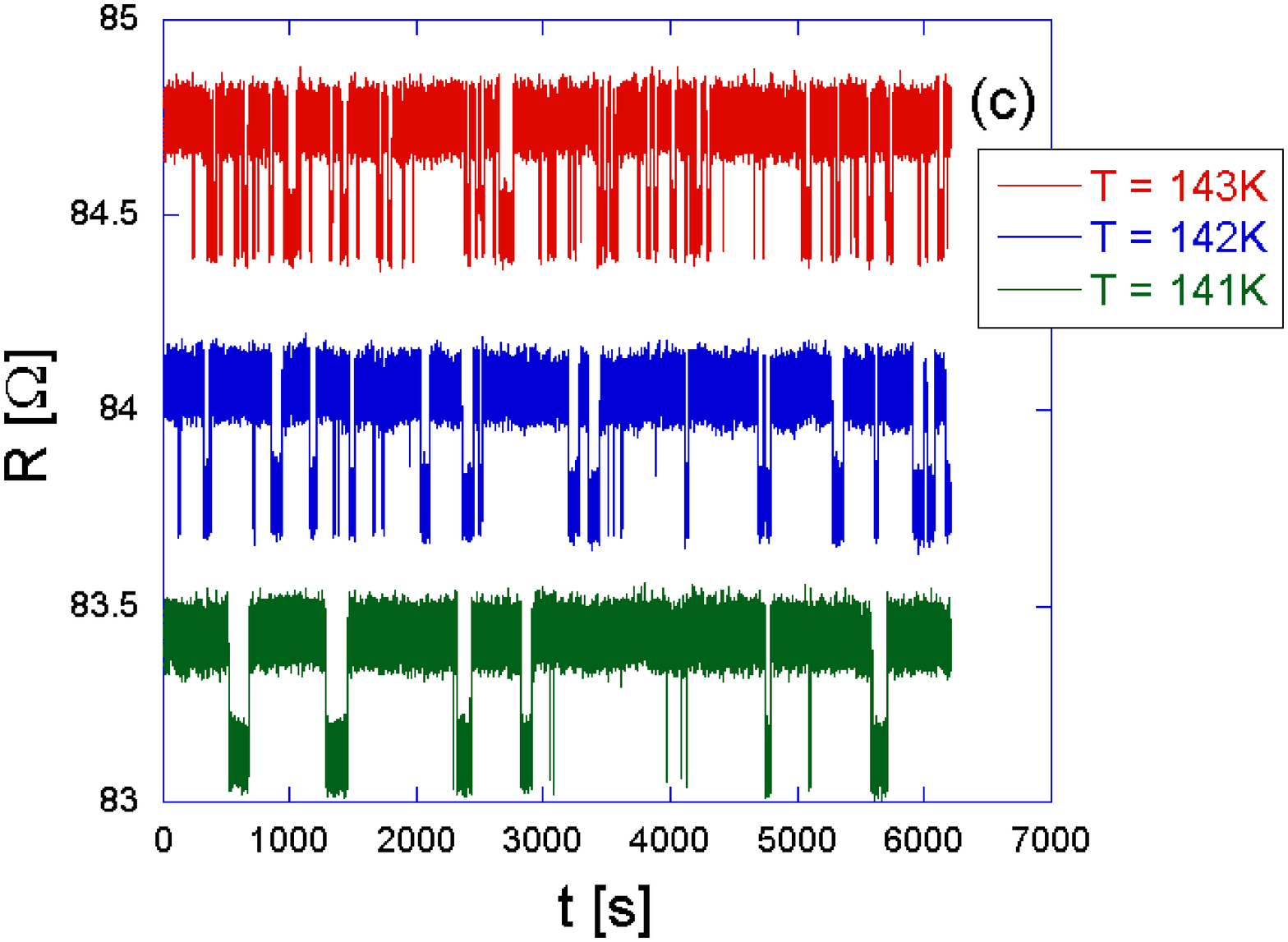}
\caption{(a) A scanning electron microscope image of a typical pattern of $\rm{ SrRuO_{3}}$. (b) $R$ vs $T$ when the pattern is fully magnetized up (red square) and down (blue circle) at zero applied magnetic field. (c) R vs t at several temperatures.}\label{FIG1}
\end{figure}

\begin{figure*}[t]
        \begin{subfigure}[a]{0.4\textwidth}
                \centering

               \includegraphics[trim=0 170 0 0,scale=0.47,angle=-90]{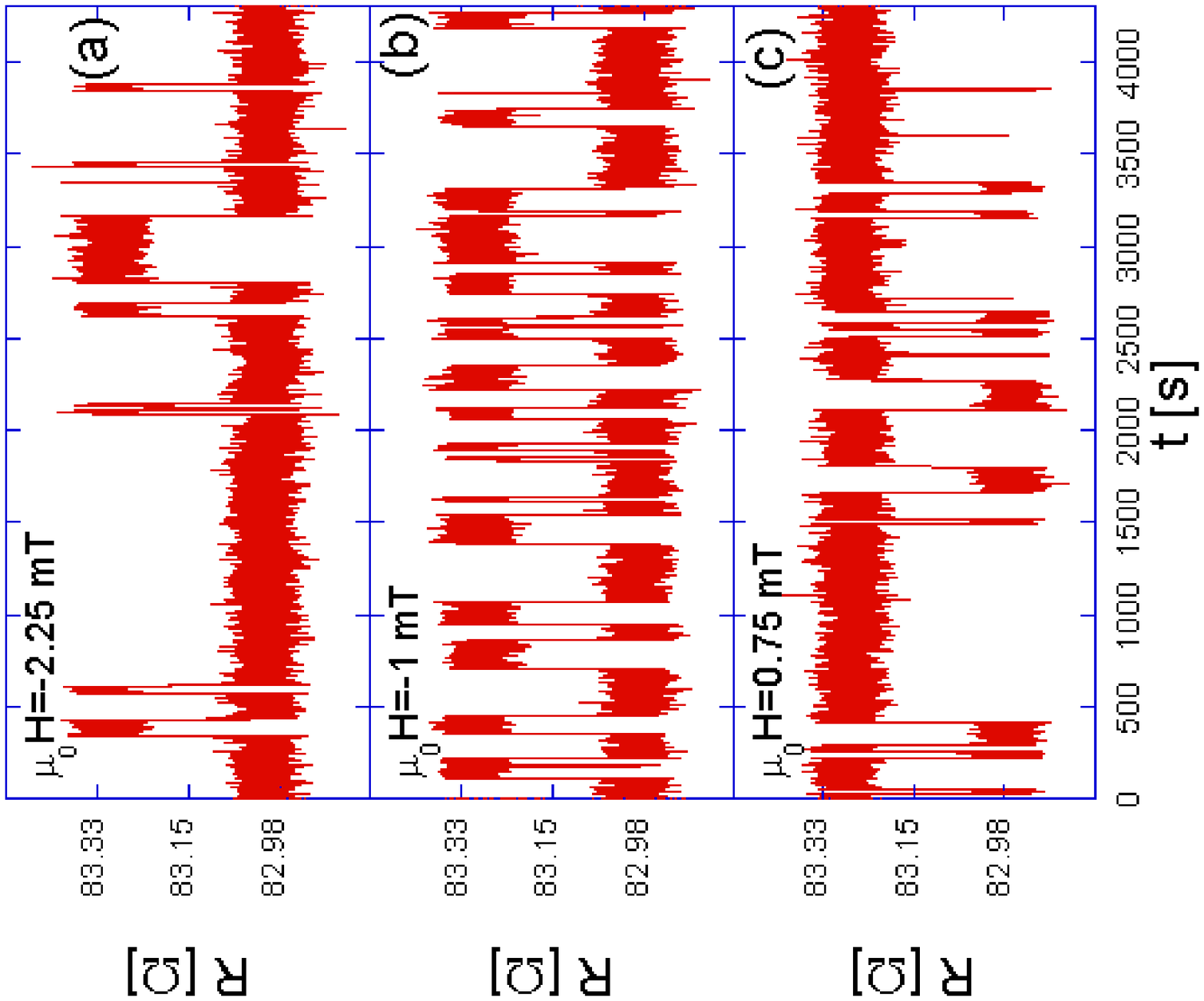}
                \label{fig:RVsTimelow}
        \end{subfigure}%
            \begin{subfigure}[b]{0.4\textwidth}
                \centering
                \includegraphics[trim=20 380 0 0,scale=0.45]{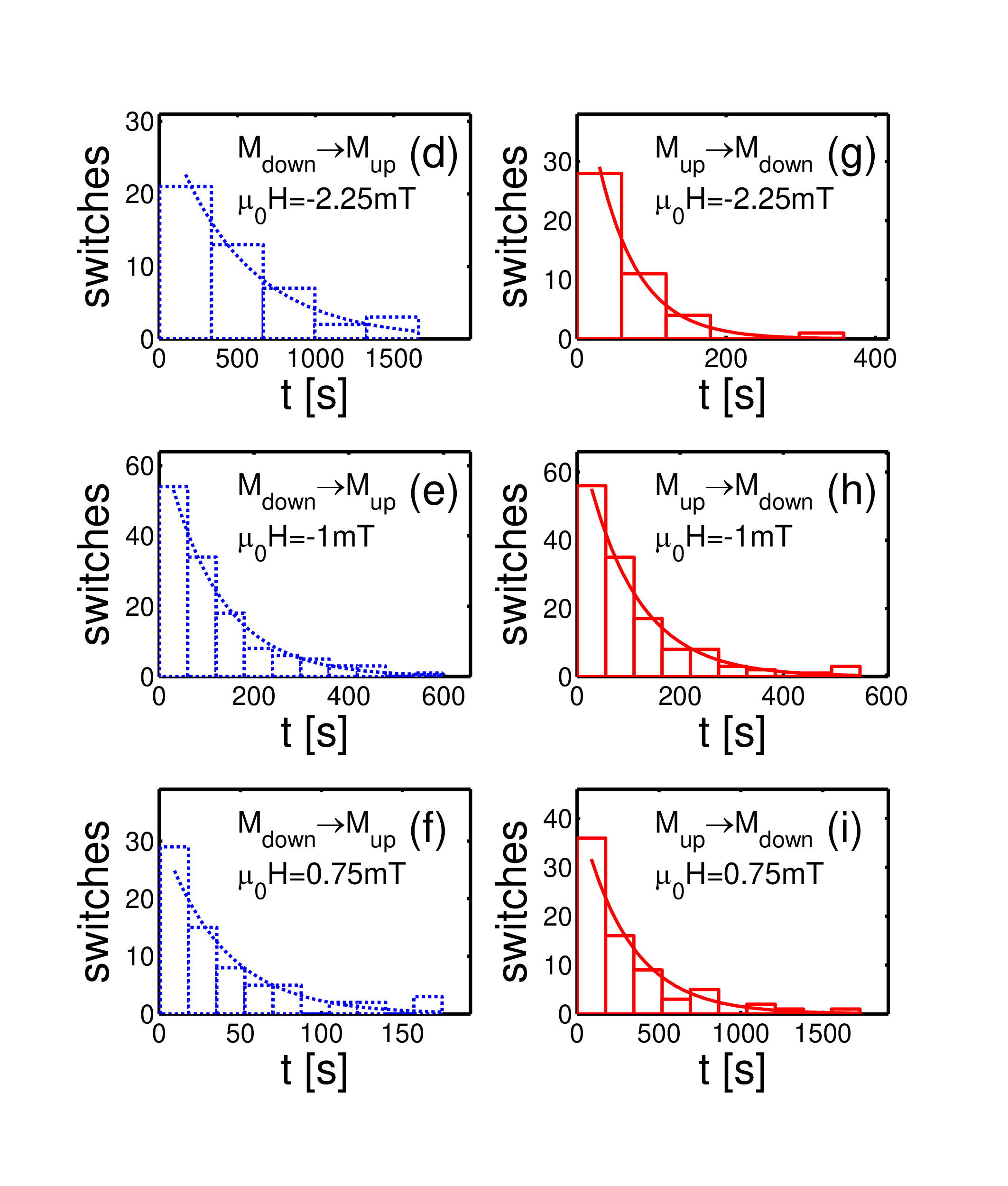}
                 \label{fig:histograms}
        \end{subfigure}

\caption{(a)–(c) $R$ vs $t$ at $T=141$ K with different applied magnetic fields($-2.25, -1$ and $ 0.75$ mT). (d)–(i) The histograms of the waiting time between reversals for the two initial states $M_{down}\protect\rightarrow M_{up}$ (dashed blue) and $M_{up}\protect\rightarrow M_{down}$ (solid red) for the same fields.}\label{RawDataHistograms}
\end{figure*}

\begin{figure}[t]
\includegraphics[scale=0.35,angle=-90]{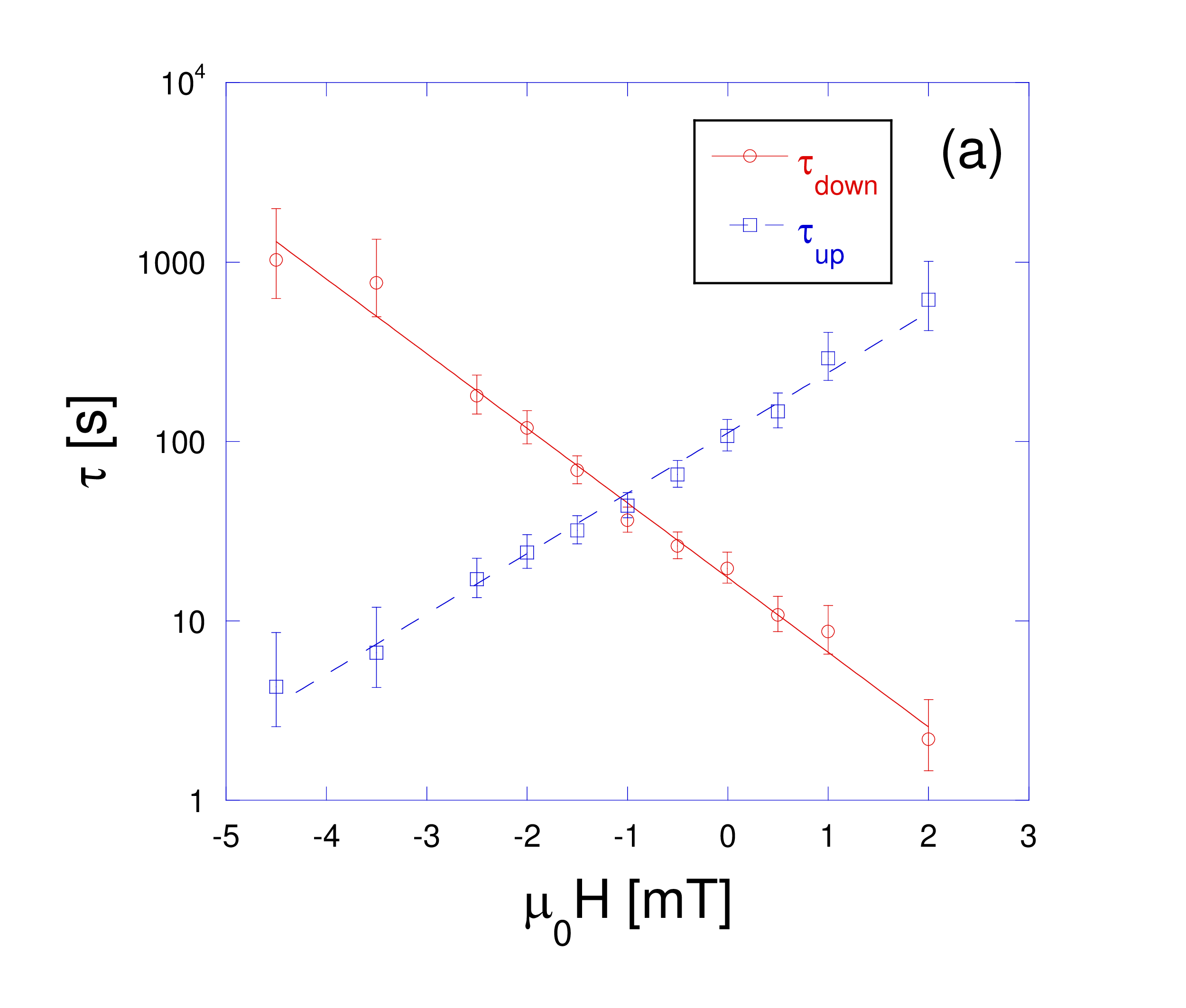}
\includegraphics[scale=0.35,angle=-90]{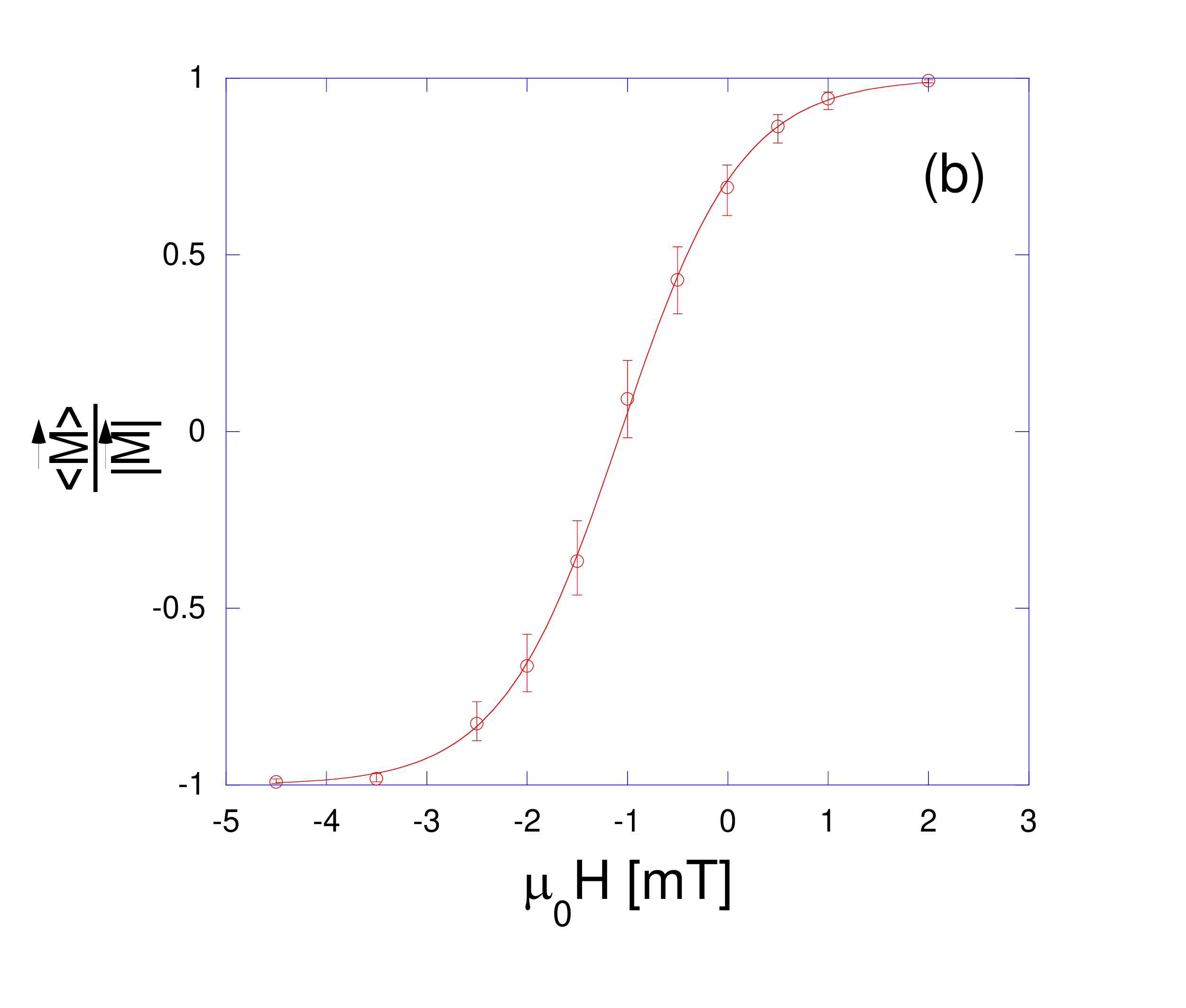}
\caption{(a) ${{\tau_{up}}}$ (blue square) and ${{\tau_{down}}}$ (red circle) as a function of the magnetic field where $T=141.5$ K. The fit is to ${{\tau}}=\alpha \exp(\gamma\mu_{0}H)$. (b) ${\langle \protect\overrightarrow{M} \rangle}/{|{M}|}$ as a function of the magnetic field. The fit is to Langevin equation ${\langle \protect\overrightarrow{M} \rangle}/{|{M}|}=\tanh(A(\mu_{0}H+B))$. The error
bars indicate a confidence interval of 95\%.
}\label{MVsH}
\end{figure}

\begin{figure}[t]
\includegraphics[trim=0 0 0 0,scale=0.35,angle=-90]{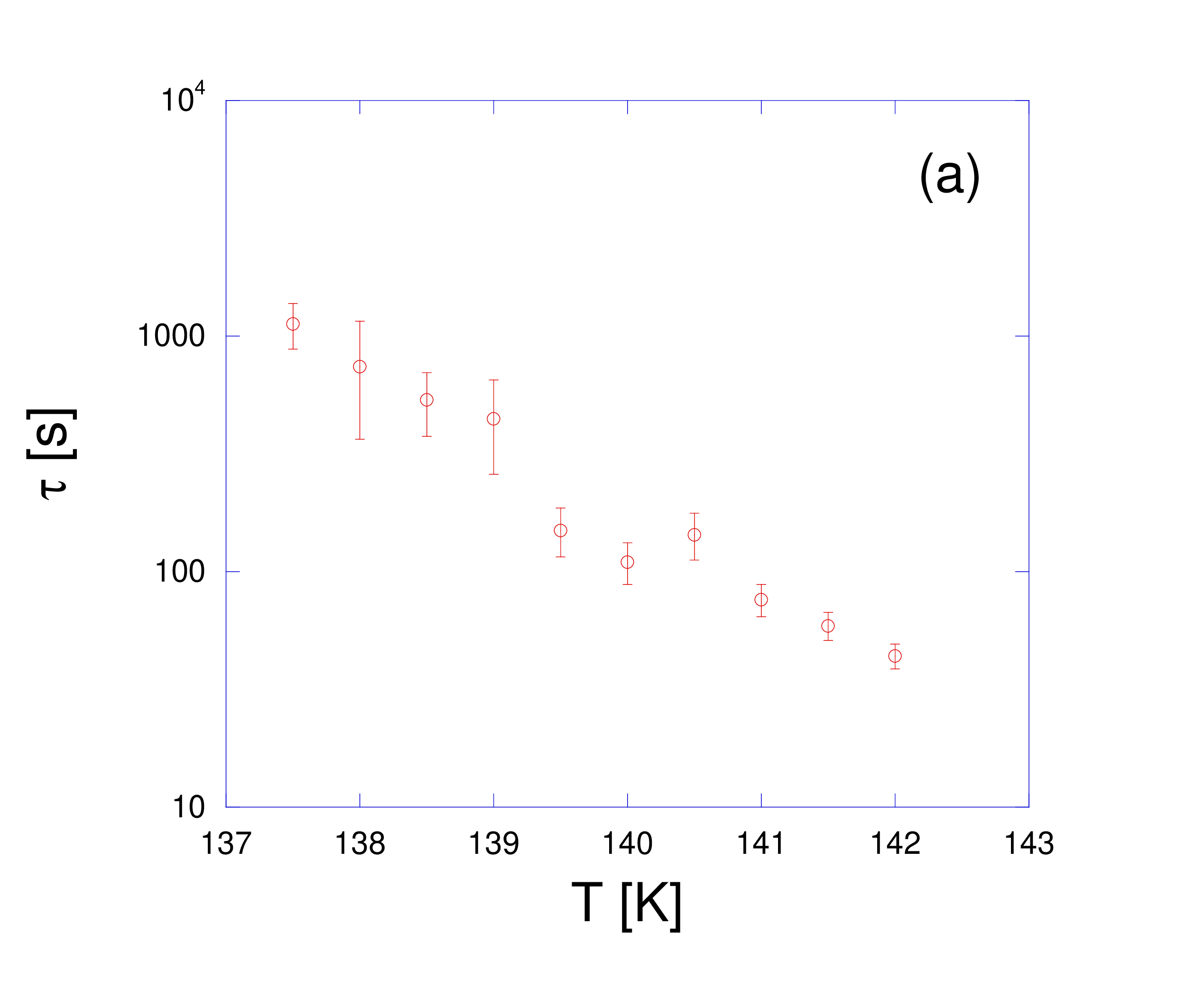}
\includegraphics[trim=0 0 0 0,scale=0.35,angle=-90]{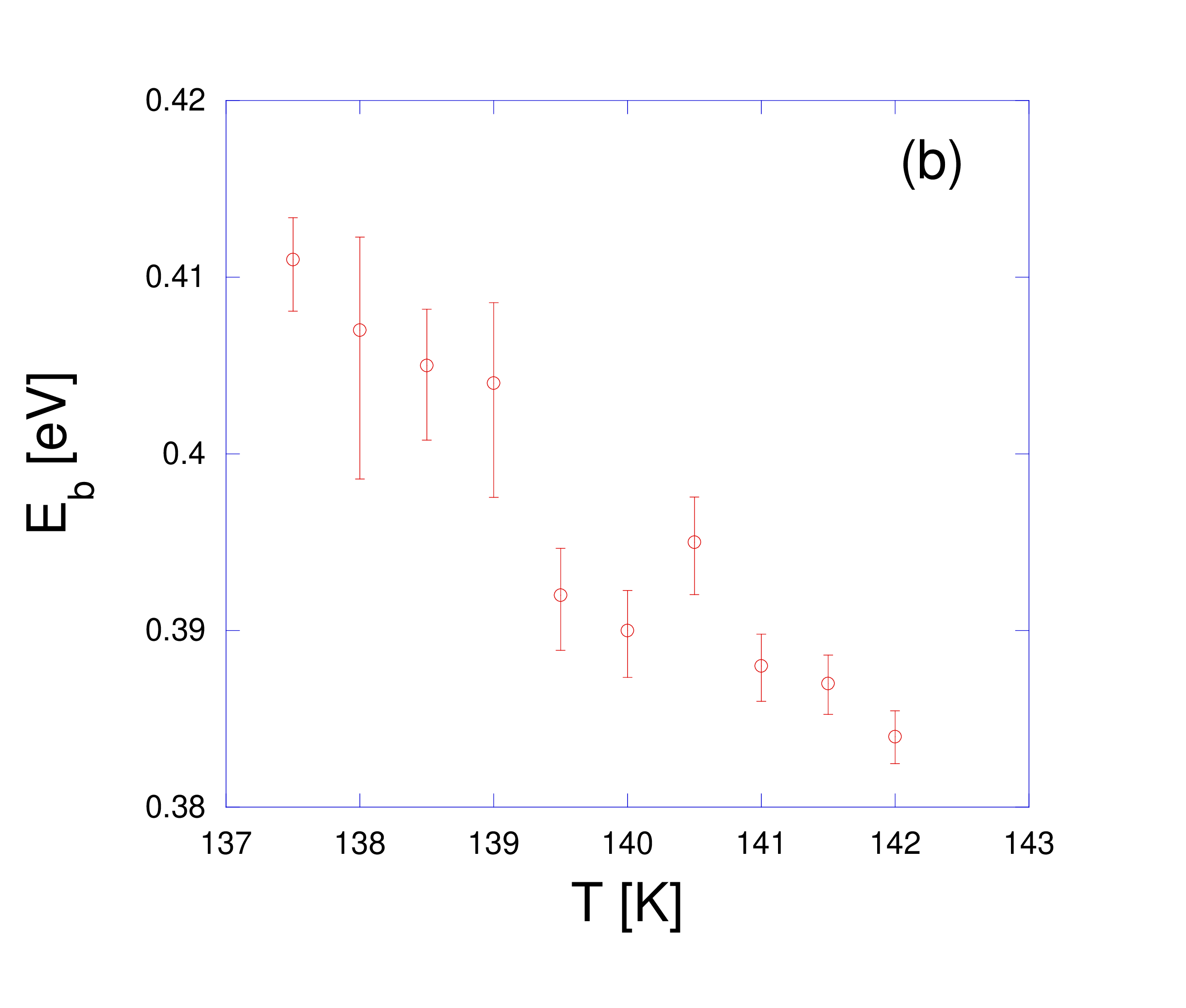}
\caption{(a) ${{\tau}}$ vs $T$. (b) $E_b$ vs $T$. The error
bars indicate a confidence interval of 95\%.}\label{EbVsT}
\end{figure}

The magnetization of ferromagnetic nanoparticles commonly exhibit thermally induced fluctuations known as a superparamagnetic behavior at a temperature interval below the Curie temperature. Superparamagnetism has been known for decades \cite{SP_Review}; however, the interest in this fundamental phenomenon has increased in recent years in connection with a wider use of spintronic devices consisting of nanoscale magnetic components \cite{NanoSpin}. Although the best way to study superparamagnetism is by exploring the superparamagnetic behavior of an individual nanoparticle, so far due to technical challenges the study of superparamagnetism  has been mainly performed with ensembles of magnetic nanoparticles where the fluctuations are not observed directly but inferred from the field and temperature dependence of the average magnetization of the ensemble \cite{LangevinMagneticPropertiesofGraphiticallyEncapsulatedNickelNanocrystals,
LangevinSuperparamagneticPropertiesOfNickelNanoparticlesInAnIonExchangePolymerFilm,
LangevinHighCoercivityAndSuperparamagneticBehaviorOfNanocrystallineIronParticlesInAluminaMatrix,
LangevinSupportedSuperparamagneticPdCoAlloyNanoparticlesPreparedFromASilicaCyanogelCoGel,
LangevinPRB2013,
LangevinJMMM2013}.

The magnetic fluctuations of an individual superparamagnetic nanoparticle are described in the framework of the N\'{e}el-Brown model \cite{Neel,Brown_pr,Brown_euro}. In its simplest form, the model describes a thermally activated process of coherent rotation of a single magnetic domain particle with uniaxial magnetic anisotropy at a temperature $T$ over an energy barrier $E_b$, and it predicts an average waiting time for reversal $\tau$ given by  $\tau= \tau_0 e^{E_b/k_B T}$, where $\tau_0$ is a sample specific constant linked to Larmor frequency with a typical value on the order of $10^{-9}$ s \cite{Cullity}. The temperature below which the waiting time exceeds the relevant measuring time (commonly on the order of 100 s) is defined as the blocking temperature $T_{b}$ given  by ${T_{b}=25K_{u}V/k_{B}}$, where ${K_{u}}$, ${V}$, and ${k_{B}}$ are the anisotropy constant, the volume of the sample, and Boltzmann constant, respectively.

The field dependence of the average magnetization $\langle\overrightarrow{M}\rangle$ is described by the Langevin equation ${\langle\overrightarrow{M}\rangle}/{|\overrightarrow{M}|}=\tanh(\mu_{0} \overrightarrow{H}\cdot\overrightarrow{M}V/k_{B}T)$, where ${\mu_{0}\overrightarrow{H}}$ is the magnetic field.
The application of the Langevin equation to describe the magnetization curves of ensembles of nanoparticles is not straightforward due to variations in the volume and shape of the nanoparticles. Therefore, any fit requires making assumptions regarding the volume distribution \cite{MQT_theory_BOOK_EnsamblesOfSmallNanoParticles}. On the other hand, in the few reports where superparamagnetic fluctuations of individual superparamagnetic nanoparticles were monitored \cite{ExperimentalEvidence,iron,HittAndCurrentSingleParticale}, the applicability of the Langevin equation was not examined.

Here we monitor superparamagnetic fluctuations in nanostructures of ${\rm SrRuO_3}$ as a function of magnetic field at different temperatures. We find that the average magnetization of an individual volume in a nanostructure (monitored by measuring the anomalous Hall effect) follows the Langevin equation, and that the volume extracted from the fit corresponds well with the actual volume in which the magnetization fluctuates. On the other hand, the rate of the fluctuations suggests a volume smaller by more than an order of magnitude. We
suggest that the switching occurs via nucleation and propagation and that the rate of the fluctuations is determined by the nucleation volume, while the volume relevant for the Langevin equation is the total volume in which the magnetization fluctuates. Due to this reversal mechanism, the blocking temperature is related to the smaller nucleation volume, which increases the superparamagnetic temperature interval by more than an order of magnitude compared to the expected interval in the case of coherent rotation. We note that there are compelling indications for magnetization reversal via nucleation and propagation even in superparamagnetic nanoparticles consisting of less than 100 atoms \cite{iron}. Therefore, we expect that the relevance of our observations to the field of superparamagnetism would  be quite general.

For this study, we use high quality epitaxial thin films of the itinerant ferromagnet ${\rm SrRuO_3}$  ($T_c{\rm \sim 150}$ K) \cite{srruo}, grown on a slightly miscut ${\rm SrTiO_3}$ substrate ($\sim 2^\circ$) by molecular beam epitaxy. The films are orthorhombic with lattice parameters $a ={\rm 5.53\ \AA}$, $b={\rm 5.57\ \AA}$, $c={\rm 7.82\ \AA}$ and they grow untwinned with the $c$ axis in the film plane and the $a$ and $b$ axes at $45^\circ$ relative to the film normal \cite{Ms}.
The films have large uniaxial magnetocrystalline anisotropy (the low-temperature anisotropy constant is $K_u\sim\rm 767 $ $\rm{{kJ}/{m^{3}}}$ corresponding to an anisotropy field higher than 7 T) \cite{Kerr} and the easy axis is
in the (001) plane. Above $T_c$, the easy axis is along $b$ (Ref. \cite{EA_Kats}) and below $T_c$, there is a reorientation transition  and the direction of the easy axis changes in the $(001)$ plane towards the film normal at a  practically constant rate of $~0.1^\circ$ per degree \cite{Ms}. The low-temperature saturation magnetization of the films is $M_s\sim213$ $\rm{{kA}/{m}}$ [corresponding to $\sim1.4$ $\rm{\mu_B}$ per ${R_u}$ (Ref. \cite{srruo})] which yields a demagnetization field that does not exceed $\sim0.2$ T, which is negligible relative to the magnetocrystalline anisotropy field. When the films are zero-field cooled, a stripe-domain structure emerges with domain walls parallel to the in-plane projection of the easy axis. The width of the magnetic domains is ${\rm \sim 200 \ nm}$ \cite{Domain strips}, and the estimated wall width is $\sim3$ nm \cite{wall_width}.

Figure 1(a) shows a typical pattern of a 7-nm-thick film which exhibits a superparamagnetic behavior. It consists of an internal rectangle ${\rm 230\pm30 \ nm \times 130\pm30 \ nm}$ connected by four narrow leads which are $ {\rm 80\pm50 \ nm}$ wide. The internal square and the leads are both made of ${\rm SrRuO_3}$. The patterns are fabricated
using a CABL-9000C e-beam high resolution lithography system (CRESTEC) followed by Ar$^+$  ion milling. The average magnetization in the internal square of the patterns is monitored by measuring the anomalous Hall effect (AHE), which is proportional to the average film-perpendicular component of the magnetization and therefore is commonly used for probing the magnetization  in patterned films. The AHE contributes to the resistance $R$ measured by driving an electrical current between contacts 1 and 3 ($I_{13}$) and measuring the voltage between contacts 2 and 4 ($V_{24})$, namely, $R=V_{24}/I_{13}$.  The measurements are performed using a PPMS system (Quantum Design) integrated with external electronics. In the measurements described here the external magnetic field is perpendicular to the film plane.

Figure 1(b) shows $R$ vs $T$ for the pattern presented in Fig. 1(a) in two states: fully magnetized up ($R_{up}$, red square) and fully magnetized down ($R_{down}$, blue circle). The resistance was measured in a range of magnetic fields (100-300 mT) to suppress superparamagnetic fluctuations and the presented values are the zero-field extrapolation of the resistance. Figure 1(c) demonstrates the superparamagnetic behavior of the pattern. The sample is cooled in zero nominal field (as noted below, there is a remanent field of $\sim 1$ mT) from above $T_c$ and the resistance  $R$ is measured as a function of time at several temperatures. The  fluctuations in  $R$ reflect a fluctuating magnetization in an area between the leads. As expected, the rate of the fluctuations decreases with decreasing temperature. The fluctuations are not between two fully magnetized states; however, the ratio between the amplitude of the fluctuations and $R_{up}-R_{down}$  is $\sim$0.6 for all temperatures indicating that the volume in which the magnetization fluctuates is the same. We note that the same qualitative behavior was observed in other samples with different ratios.

If the probability to switch between states is time-independent, we expect that the probability to wait a time $t$  for switching will be described by an exponential distribution $\rho(t)=(1/{\tau})\exp(-t/{\tau})$ where $\tau$ is the average of $t$. Figures 2(a)-2(c) show typical time dependence measurements of $R$ at 141 K for different applied magnetic fields (-2.25, -1, and 0.75 mT). The total measuring time for each field is at least $400$ min. Figures 2(d)-2(i) show the histograms of $t$ for initial magnetization-up state ($M_{up}\rightarrow M_{down}$) and magnetization-down state ($M_{down}\rightarrow M_{up}$) with the mentioned different applied magnetic fields. For each histogram, the fit is to $\int^{t+\Delta/2}_{t-\Delta/2}(1/{\tau})\exp(-t/{\tau})dt$ multiplied by the total number of switches where $\Delta$ is the corresponding time window of the histogram. We note a change in the waiting time distribution for the two initial magnetic states as a function of the applied magnetic field.

  Figure 3(a) shows the dependence of $\tau$ for initial magnetization-up state ($\tau_{up}$) and magnetization-down state ($\tau_{down}$) on the applied magnetic field at $T=141.5$ K. The fit is to $\tau=\alpha\rm{exp}(\gamma\mu_{0}H)$, where $\alpha$ and $\gamma$ are fitting parameters.

Figure 3(b) shows the normalized average magnetization ${\langle\overrightarrow{M}\rangle}/{|\overrightarrow{M}|}=({\tau_{up}-\tau_{down}})/({\tau_{up}+\tau_{down}})$ as a function of the magnetic field at $T=141.5$ K. The fit is to the Langevin equation ${\langle\overrightarrow{M}\rangle}/{|\overrightarrow{M}|}=\tanh[A(\mu_{0}H+B)]$ with $A=836\pm15$ $\rm{{1}/{T}}$ and $B=1.06\pm0.015$ mT. The parameter $B$ is attributed to a small remanent field of the superconducting coil of the PPMS and the ambient field.

  The energy difference between the up and down states is $\Delta{E}=2VM\mu_{0}H \cos \theta$, where $V$ is the volume in which the magnetization fluctuates and $\theta\sim45^\circ$ is the angle between the magnetization and the magnetic field. The parameter $A$ is given by $A={VM \cos 45^\circ}/{k_{B}T}$ . Based on the observation that the average magnetization at 141.5 K is 0.27 of the saturation magnetization, the extracted $V$ from the fit is $40\,000 \pm1000 \ \rm{nm^3}$, which is $\sim 1/5$ to $1/2$ of the total volume between the leads, assuming a dead layer thickness of $\sim 2 \ {\rm nm}$ \cite{DeadLayer}. Considering the ratio between the amplitude of the fluctuations and ($R_{up}-R_{down}$), the extracted volume is reasonable.

Since we follow the waiting time for reversal of an individual volume, we can in principle extract the volume magnitude directly from the average waiting time $\tau$ at zero field given by $\tau=\tau_{0}\exp({E_{b}/k_{B} T})$, where  $\tau_{0}\sim10^{-12}$ s \cite{Kerr,OurMQT}. The energy barrier $E_b$ is given in the case of coherent rotation by $E_b=K_{u}V$, where $K_u$ is the anisotropy constant.  The value of $\tau$ at zero effective  field as a function of temperature is shown in Fig. 4(a), and the corresponding  $E_{b}$ is shown in Fig. 4(b). Assuming the volume extracted from the fit to the Langevin equation  ($40\,000 \ \rm{nm^3}$), we extract  $K_u=1.55\pm0.01$ $\rm{{kJ}/{m^{3}}}$ at 141.5 K, which would imply an anisotropy field of $\sim0.055$ T. However, applying an in-plane magnetic field [$45^\circ$ from the easy axis in the (001) plane] of 8 T reduces the AHE by less than 50\%, namely, the actual anisotropy is clearly orders of magnitude higher than $0.055$ T. The lack of full magnetization alignment with the 8 T in-plane field is consistent with the description of ${\rm SrRuO_3}$ in the framework of the anisotropic Heisenberg model \cite{EA_Kats} which would imply $K_u=56$ $\rm{{kJ}/{m^{3}}}$ and an anisotropy field of 2 T. We note that using this value of anisotropy and the total volume in which the magnetization fluctuates would yield an average waiting time of more than $10^{400}$ s. Furthermore, we would expect that the blocking temperature would be on the order of 100 mK below $T_c$. Namely, the observed superparamagnetic temperature interval is by more than an order of magnitude larger than what we would expect for coherent rotation of that volume. A plausible reason for the different extracted volumes is a scenario of nucleation and propagation where the waiting time is determined by the barrier for nucleation while the volume relevant for the Langevin equation is the total volume in which the magnetization reverses during nucleation followed by propagation. Such a nucleation-propagation scenario is consistent with the magnetization reversal behavior below the blocking temperature \cite{OurMQT} and it is also consistent with Monte Carlo simulations we have performed.

In a nucleation-propagation scenario, the relevant energy for the switching rate is the energy cost of a domain wall whose length is $L$ given  by
$E_{DW}=(DL)(4\sqrt{AK_u})$ \cite{Chikazumi}, where $A={JS^2}/{a}$ is the exchange stiffness, $D$ is the sample thickness, $K_u$ is the anisotropy constant, $J$ is the exchange constant, $S$ is the spin number, and $a$ is the length between neighboring spins. In our case, $D=5$ nm, $J=26.3K_B$, and $a=0.4$ nm. The other parameters are temperature dependent and at $T=141.5$ K their values are $E_{DW}=0.387$ eV, $S={\langle\overrightarrow{M}\rangle}/{Ms}=0.27$, and $K_u=56$ $\rm{{kJ}/{m^{3}}}$. Using these parameters we find that the wall length is about $52$ nm, consistent with the dimensions of the volume in which the magnetization fluctuates.

This scenario assumes that the intermediate state following  the nucleation is unstable  and propagation occurs in time scales much shorter than the time scale spent at each of the end states. For this to happen, it is required that the state following nucleation in ournanostructures is unfavorable energetically, which we confirmed using Object oriented micromagnetic framework (OOMMF) simulation \cite{OOMMF}.

In conclusion, using patterned nanostructures of ${\rm SrRuO_3}$ we demonstrate the applicability of the Langevin equation in describing the magnetic fluctuations in an individual volume where the volume estimated from the fit to the Langevin equation is consistent with the pattern dimensions. We find that the time intervals between reversals yield a volume smaller by more than an order of magnitude. We suggest a scenario of nucleation followed by propagation where the barrier for nucleation is determined by the energy cost of a domain wall whose length is consistent with the dimensions of the volume in which the magnetization fluctuates. The fact that the reversal process does not appear to affect the volume extracted from the fit to the Langevin equation suggests that the time spent in the metastable states of nucleation and propagation is negligible compared to the time spent at the two end states. We also note that this nucleation-propagation mode significantly extends the temperature interval where superparamagnetic behavior can be observed.

We acknowledge useful discussions with E. M. Chudnovsky. L.K. acknowledges support by the Israel Science Foundation founded by the Israel Academy of Sciences and Humanities.  J.W.R. grew
the samples at Stanford University in the laboratory of M. R. Beasley.

\end{document}